\documentclass[10pt]{article}
\usepackage[a4paper,top=3cm,bottom=3cm,left=3cm,right=3cm,marginparwidth=1.75cm]{geometry}
\usepackage[utf8x]{inputenc}
\usepackage[english]{babel}
\usepackage[T1]{fontenc}

\usepackage{sectsty}
\sectionfont{\normalsize}
\subsectionfont{\normalsize}
\subsubsectionfont{\normalsize}
\paragraphfont{\normalsize}
\usepackage{amssymb,amsmath}
\usepackage{graphics}
\usepackage{setspace}
\usepackage{float}
\usepackage{graphicx}
\usepackage[colorinlistoftodos]{todonotes}
\usepackage[hang,flushmargin]{footmisc} 
\usepackage{hyperref}
\usepackage{booktabs}
\usepackage[sort,round]{natbib}

\begin{document}

\noindent
\begin{center}
\large{\bf An Application of Rubi: Series Expansion of the Quark Mass Renormalization Group Equation}
\end{center}

\vspace{1mm}
\begin{center}
\text{Alexes Mes$^{a}$, Jed Stephens$^{b}$}
\end{center}

\begin{center}
{\it $^{(a)}$MSXALE002@myuct.ac.za,   $^{(b)}$STPJED001@myuct.ac.za}
\end{center}

\vspace{2mm}

\begin{abstract}
\noindent We highlight how Rule-based Integration (Rubi) is an enhanced method of symbolic integration which allows for the integration of many difficult integrals not accomplished by other computer algebra systems. Using Rubi, many integration techniques become tractable. Integrals are approached using step-wise simplification, hence distilling an integral (if the solution is unknown) into composite integrals which highlight yet undiscovered integration rules.
The motivating example we use is the derivation of the updated series expansion of the quark mass renormalization group equation (RGE) to five-loop order. This series provides the relation between a light quark mass in the modified minimal subtraction ($\overline{\text{MS}}$) scheme defined at some given scale, e.g. at the tau-lepton mass scale, and another chosen energy scale, $s$. This relation explicitly depicts the renormalization scheme dependence of the running quark mass on the scale parameter, $s$, and is important in accurately determining a light quark mass at a chosen scale. The five-loop QCD $\beta(a_s)$ and $\gamma(a_s)$ functions are used in this determination. 

\vspace{3mm}
\end{abstract}

\section{Extensions to CAS by Rubi}\label{sec:CASbyRubi}
Computer Algebra Systems (CAS) such as Mathematica \citep{Mathematica} and SymPy \citep{meurer2017sympy} (the popular open-source alternative implemented in Python), have built-in symbolic integral routines. 
Rule-based Integration (Rubi) developed by \citet{richrule} is principally a package (designed for Mathematica) that provides a method of symbolic integration organized by decision tree pattern matching, which matches the form of the integral against known integral rules. 
Rubi comprises 6700+ rules, collated from familiar favourites \citet{A&S, CRC, G&R} and in doing so it offers not only a means of integrating, but a growing complete reference for integration rules. These rules are in human-readable form with cross references to Rubi rule numbers, and to the source. Rubi can also print the rules applied at each stage of solving the integral -- a useful technique for pedagogical and diagnostic purposes.

Without proper consideration it may not be obvious why Rubi marks a significant improvement to effectively solving integrals.
The effectiveness of these routines have been independently investigated by \citet{Abbasi-CAIIT} with the results presented in Table \ref{tab:antiderivative-grade}. Comparing Rubi 4.15.2, Mathematica 11.3 and SymPy 1.1.1 \citet{Abbasi-CAIIT} divides the quality of integral's antiderivatives into four groups.\textit{Group A} consists of integrals that were easily solved: where the antiderivative is optimal in quality and leafsize. \textit{Group B} is the group of integrals which were solved, but the leafsize twice that of optimal. \textit{Group C's} integrals were solved, but the solution contains hypergeometric functions, special functions or imaginary units while the optimal antiderivative does not. Finally \textit{Group F} are all integrals which cannot be solved by the CAS. See \citet{Abbasi-CAIIT} for more details. 

\begin{table}[ht]
\begin{center}
\label{tab:antiderivative-grade}
\caption{Antiderivative Grade distribution for each CAS}
\begin{tabular}{@{}ccccc@{}}
\hline
System      & \% A grade & \% B grade & \% C grade & \% F grade \\ \hline
Rubi 4.15.2       & 99.76      & 0.08       & 0.06       & 0.1        \\
Mathematica 11.3 & 75.37      & 8.46       & 15.81      & 2.67       \\
SymPy 1.1.1     & 30.29      & 0          & 0          & 69.71      \\ \hline
\footnotesize Adapted: \cite{Abbasi-CAIIT} pg. 6
\end{tabular}
\end{center}
\end{table}

Rule-based integration is the focus of much attention in development not only by \citet{richrule}, but by others.
For example SymPy 1.1.1 currently fairs comparatively poorly in symbolic integration to other CAS (Table \ref{tab:antiderivative-grade}). However, the rules and implementation (pattern matching in a decision tree) behind Rubi are currently being developed into SymPy, see \citet{sympy-rubi} for details.
This would clearly improve the quality of this open source alternative.

\section{Motivating Example: Quark Mass Renormalization Group Equation}

Having examined how powerful Rubi is as a symbolic integration tool, we now explore how it can be applied in computation, using the quark mass renormalization group equation as a motivating example.
This example is purposefully chosen, as will become apparent later, because the integral of interest is challenging for CAS.

Along with the strong coupling, the quark masses are fundamental parameters of Quantum Chromodynamics and it is therefore important to accurately know their numerical values. Further, it is important to know the scale dependence of these values. 

In QCD, as in Quantum Electrodynamics (QED), one removes the present divergences with a technique known as renormalization. A nonphysical renormalization scale parameter, $\mu$, is introduced in the renormalization procedure to represent the point at which one performs the subtraction of the divergences to render the amplitudes finite. Both the renormalized coupling $\alpha_s(\mu^2)$, and the quark masses, $\overline{m}_q(\mu^2)$, depend on the renormalization scheme used to define the theory, and on the scale parameter, $\mu$. If we set $\mu$ approximately equal to the scale of the momentum transfer Q in a particular interaction,  $\alpha_s(\mu^2 \approx Q^2)$ becomes the effective strength of the strong coupling for that interaction \citep{tanabashi2018m}. Throughout this paper, we make use of the physical energy scale parameter, $s$ (where $s\,=\,Q^2$). The scale dependence of $\alpha_s(s)$ and $\overline{m}_q(s)$ is governed by corresponding renormalization group equations (RG equations) which rely on QCD's anomalous dimensions as input.

The strong coupling, $\alpha_s(s)$, satisfies the differential RGE \citep{davier2006physics}:
\begin{equation} \label{eq:A.1}
\frac{d a_s}{d \ln s} = \beta(a_s) = -a_s^2 \, (\beta_0 \, + \,  a_s\, \beta_1\, + \,  a_s^2\, \beta_2 \, + \, a_s^3\, \beta_3 \, + a_s^4\, \beta_4)
\end{equation}

where the $\beta(a_s)$ function is known up to $\mathcal{O}(a_s^6)$, and $a_s \, \equiv \, \frac{\alpha_s}{\pi} \, = \,  \frac{g_s^2}{4\pi^2}$ ($g_s$ is the gauge coupling of QCD). Given the renormalization point, the $\beta(a_s)$ function describes how the strong coupling depends on the momentum transfer. 
 
The quark masses, $m_q(s)$, satisfy the differential RGE \citep{davier2006physics}:
\begin{equation} \label{eq:A.2}
\frac{1}{\overline{m}_q} \frac{d \overline{m}_q}{d \ln s} = \gamma(a_s) = -a_s\, (\gamma_0\, + \,  a_s\, \gamma_1 \, + \, a_s^2\, \gamma_2 \,  + \, a_s^3\, \gamma_3 \,  + \, a_s^4\, \gamma_4)
\end{equation}

where the $\gamma(a_s)$ function is an anomalous dimension and is known up to $\mathcal{O}(a_s^5)$. The $s$-dependence of $a_s$ and $\overline{m}_q$ in Eqs.(\ref{eq:A.1})-(\ref{eq:A.2}) is implicit i.e. $a_s = a_s(s)$ and $\overline{m}_q = \overline{m}_q(s)$. 

The coefficients of the
$\beta(a_s)$ function, which are now known to five-loop order \citep{baikov2017five, luthe2017complete, herzog2017five}, are given (for three active quark flavours) by: $\beta_0 = 9/4$, $\beta_1 = 4$, etc. While the $\gamma(a_s)$ function coefficients, also currently known to five-loop order \citep{baikov2014quark, chetyrkin1997quark, vermaseren19974}, are: $\gamma_0 = 1$, $\gamma_1 = 91/24$, etc., for three flavours. Throughout this paper we work in the modified minimal subtraction scheme ($\overline{\text{MS}}$) \citep{veltman1972regularization, bardeen1978deep}. This renormalization scheme is the most commonly used scheme in QCD perturbation theory.

It is important to be aware that there are consequences to crossing flavour thresholds. We concentrate on deriving the series expansion of Eq.(\ref{eq:A.2}) in the light quark sector (the up-, down- and strange-quark). If we proceed to higher energies (into the heavy quark region), the renormalization scale crosses quark mass flavour thresholds: finite threshold corrections appear \citep{chetyrkin1998decoupling} and the scale dependence of the mass then needs to be matched above and below the threshold. One has to specify a new initial condition for the running coupling constant at each threshold. In principle, Eq.(\ref{eq:A.8}) is valid for any number of quarks $n_f$, between two thresholds provided the correct initial values are used. For the light quark sector we set $n_f =3$.

The recent calculation of the $\beta(a_s)$ function to five-loop order by \citet{baikov2017five, luthe2017complete, herzog2017five}, has ensured that the series expansion of the running quark mass can now be calculated to five-loop order. Previously this series expansion has been calculated by \citet{chetyrkin1997estimations} to four-loop order, which built on the three-loop order calculation done by \citet{kniehl1996dependence}. \citet{chishtie2018systematic} provide a recent exposition into the topic to four-loop order without the use of CAS, but stop short of explicitly providing the series expansion. The perturbative series solution to Eq.(\ref{eq:A.2}) involves performing a Taylor expansion of $\overline{m}_q(s)$ at some reference scale $s=s^*$, in powers of $\eta \, = \, \ln(s/s^*)$. To the third- and fourth-loop this calculation is a fairly trivial exercise. At higher loop orders, however, this computation becomes more difficult and CAS such as Mathematica and SymPy struggle to intuitively solve the RG equation without the additional use of Rubi. We outline the method for using Rubi to find a perturbative solution to Eq.(\ref{eq:A.2}) in the following section. The derivation is purely symbolic.

\section{The Perturbative Series Expansion of \texorpdfstring{$\overline{m}_q(s)$}{mq(s)}} \label{sec:seriesexpansion}

The quark mass RG equation (Eq.(\ref{eq:A.2})) can be identified as a linearly separable differential equation. As such, we are able to exactly solve for $\overline{m}_q$ given the coefficients of the $\beta(a_s)$ and $\gamma(a_s)$ functions to a certain order. The exact solutions to leading and next-to-leading order are given in \cite{kniehl1996dependence}. However, it is difficult to obtain the exact solution of $\overline{m}_q$ at higher orders, and this becomes a numerical procedure. Therefore, it is more lucid to solve the renormalization group equations in terms of a power expansion; since this type of solution provides insight into the renormalization scheme dependence of the running quark mass on the energy scale parameter $s$, at higher powers. This is important in accurately determining the light quark mass at a chosen scale. Hence, we proceed with determining a perturbative series expansion of Eq.(\ref{eq:A.2}).

This is achieved by dividing Eq.(\ref{eq:A.2}) by Eq.(\ref{eq:A.1}) and linearly separating the differentials to yield
\begin{equation} \label{eq:A.3}
\frac{d \overline{m}_q}{\overline{m}_q} = \frac{\gamma(a_s)}{\beta(a_s)} \,d a_s 
\end{equation}

where $\gamma(a_s)$ and $\beta(a_s)$ were defined in Eqs.(\ref{eq:A.1})-(\ref{eq:A.2}).

Integrating Eq.(\ref{eq:A.3}) leads to 
\begin{equation} \label{eq:A.4}
\ln \left( \frac{\overline{m}_q(s)}{\overline{m}_q(s^*)} \right) = \int_{a_s(s^*)}^{a_s(s)} \, da_s' \frac{\gamma(a'_s)}{\beta(a'_s)}
\end{equation}

Which can be easily rearranged to find
\begin{equation} \label{eq:A.5}
\overline{m}_q(s) = \overline{m}_q(s^*) \, \exp \left(\int_{a_s(s_0)}^{a_s(s)}  da_s' \, \frac{\gamma(a'_s)}{\beta(a'_s)} \right)
\end{equation}
where $\bar{m_q}(s^*)$ is the initial condition.

Both Mathematica and Rubi can be used in attempts to solve the integral in Eq.(\ref{eq:A.5}). 
What is of interest is how each of these CAS approach solving the chosen problem. 
In terms of the integral classification we introduced in Section \ref{sec:CASbyRubi}, we can classify the integral in Eq.(\ref{eq:A.5}) as a \textit{Group F} integral, which means that Rubi and Mathematica are unable to solve the integral analytically. 
Naively using Mathematica's inbuilt integration function immediately yields an answer in terms of a \textsf{RootSum} object. 
Mathematica then struggles to find the definite integral (and series expansion) due to infinities arising from the logarithmic terms in this \textsf{RootSum} object. 
Mathematica's solution has a low interpretability and it's not what part of the integration process eventually yields the \textsf{RootSum} object.

Comparatively Rubi's attempt at the integral is a partial solution involving lower order integrals.
The key advantage Rubi offers here is in simplification and clarity in identifying the unevaluated sections of the problem. 
Rubi performs the integral step-wise while printing the integration rule that it employs at each stage -- which is is worth emphasising.
This allows the researcher to focus on what Rubi does \textit{not} know.
Should the researcher find an analytical solution for these unknown integrals it is easy to develop the appropriate rule and submit it to the Rubi GitHub project.
The high interpretability of Rubi's attempt means that these remaining integrals can then be suitably approximated. 
Finding the series expansion from this point is straightforward. 

Rubi's attempt at the indefinite version of the integral in Eq.(\ref{eq:A.5}) yields

\begin{flalign} 
& \begin{aligned} \label{eq:A.6}
F({a'_s}) \, & = \, \int da_s' \, \frac{\gamma({a'_s})}{\beta({a'_s})} \\
& = \,  \dfrac{\gamma_0 \ln({a'_s})}{\beta_0}  - \dfrac{1}{4\,\beta_0\,\beta_4} \, \Bigg\{ \big(\beta_4\,\gamma_0 - \beta_0\,\gamma_4\big)\, \ln\big(\beta_0 + \beta_1\,{a'_s} + \beta_2\,{a'_s}^2 + \beta_3\, {a'_s}^3 + \beta_4\, {a'_s}^4\big)\\[.1cm]
 & \, + \,  I_0 \, \big(3\,\beta_1\,\beta_4\,\gamma_0 - 4\beta_0 \,\beta_4\, \gamma_1 + \beta_0\,\beta_1\,\gamma_4\big)\, + \, 2 \, I_1 \, \big(\beta_2\,\beta_4\,\gamma_0 - 2\,\beta_0\,\beta_4\,\gamma_2 + \beta_0\,\beta_2\,\gamma_4\big) \\[.1cm]
 &+ I_2 \, \big(\beta_3\,\beta_4\,\gamma_0 - 4\,\beta_0\,\beta_4\,\gamma_3 + 3\,\beta_0 \,\beta_3\,\gamma_4\big) \Bigg\}
 \end{aligned} &
\end{flalign}

where \begin{equation} \label{eq:A.7}
I_n = \int d{a'_s} \, \frac{a_{s}'^{n}}{\beta_0 \,+\, \beta_1\,{a'_s} \,+\, \beta_2\,{a'_s}^2 \,+\, \beta_3\,{a'_s}^3 \,+\,\beta_4\,{a'_s}^4} 
\end{equation}

The integrals $I_n$, do not at present have an analytic solution in terms of algebraic functions -- at least, they are unknown by Rubi. At this stage, however, Mathematica is able to re-write these integrals in terms of \textsf{RootSum} objects (without logarithmic divergences) that can be suitably simplified when the series expansion is performed. 

The definite integral of Eq.(\ref{eq:A.6}) is found simply by using the Fundamental Theorem of Calculus\footnote{An assumption of the Fundamental Theorem of Calculus is that the function to be integrated must be continuous. In the present case, the integrand is a rational function and therefore continuous up to isolated poles in the complex plane.}. The upper bound of the definite integral, the scale dependent strong coupling $a_s(s)$, is rewritten as its perturbative solution in terms of some known $a_s(s^*)$ (e.g. at the tau-lepton mass scale) up to $\mathcal{O}(a_s^6)$ \citep{davier2006physics}. The resulting definite integral is quite lengthy, despite some simplification occurring between polynomial sums arising from the $I_n$ integrals in Eq.(\ref{eq:A.6}). It can be viewed in the supplementary Mathematica notebook.

Finally focusing on Eq.(\ref{eq:A.5}), we exponentiate the definite integral, and perform a series expansion at some reference scale $s\,=\, s^*$. 

Reordering the perturbative solution in terms of $a_s(s^*)$ yields
\begin{flalign} 
& \begin{aligned} \label{eq:A.8}
\overline{m}_q(s) \,& = \, \overline{m}_q(s^*) \, \Bigg\{1 - a(s^*) \,\gamma_ 0 \,\eta + \frac{1}{2} \, a^2(s^*)\, \eta\, \Big[-2 \,\gamma_ 1 + \gamma_ 0\, (\beta_0 \,+ \,\gamma_ 0)\, \eta\Big]\\[.1cm]
&- \frac{1}{6} \, a^3(s^*) \,\eta \,\Big[6 \,\gamma_ 2 - 3 \,\Big(\beta_ 1\, \gamma_ 0\, + 2 \,(\beta_0 \,+\,\gamma_ 0) \, \gamma_ 1\Big)^{} \, \eta\, +\, \gamma_ 0 \,(2 \,\beta_0^2 \,+ 3 \,\beta_0 \,\gamma_ 0 \,+\, \gamma_0^2) \,\eta^2 \Big] \\[.1cm]
&+ \frac{1}{24}\,  a^4(s^*) \, \eta \, \Big[-24\, \gamma_3 \,+ \,12 (\beta_ 2\,
 \gamma_ 0\, +\, 2 \beta_ 1\,\gamma_ 1 \,+ \,\gamma_ 1^2 \,+ 3\, \beta_0\, \gamma_ 2 \,+ 2\, \gamma_ 0 \,\gamma_ 2)\, \eta\,\\[.1cm]
& - \, 4\, \Big(6\, \beta_0^2\, \gamma_ 1 \,+ \, 3\, \gamma_ 0^2\, (\beta_ 1 \,+ \,\gamma_1) \,+\, \beta_0\, \gamma_ 0 \,(5 \,\beta_ 1\, + \,9\, \gamma_ 1)\Big) \,\eta^2 \,+ \,\gamma_ 0 \,(6\, \beta_0^3\, +\, 11\,\beta_0^2 \,\gamma_ 0 \\[.2cm]
&+ 6\, \beta_0\, \gamma_ 0^2\, +\, \gamma_ 0^3) \,\eta^3 \Big] \\[.1cm]
&+ \dfrac{1}{120}\, a^5(s^*)\, \eta \,\Big[-120\, \gamma_ 4 \,+\,  \dfrac{1}{\beta_0} 60\, \Big(-7 \,\beta_ 1 \,\beta_ 2\, \gamma_0 \,+\, 4 \,\beta_0^2\, \gamma_ 3\, +\, \beta_0\,(7 \,\beta_ 1 \,\gamma_ 0 \,+ \,\beta_ 3 \,\gamma_ 0\,\\[.1cm]
&+ 2\, \beta_ 2\, \gamma_ 1 \,+ \,3\,\beta_ 1 \,\gamma_ 2\, +\,  2 \gamma_ 1 \,\gamma_ 2 
\, +\, 2\,\gamma_ 0 \,\gamma_ 3)\Big)\, \eta \,- \,20\,\Big(3 \beta_ 1^2 \,\gamma_ 0 \,+\, \beta_ 1\, (14 \, \beta_0 \,+\,
9 \,\gamma_ 0)\, \gamma_ 1 \,\\[.1cm]
&+ 3\, (2 \beta_0\, +\, \gamma_ 0) (\beta_ 2\, \gamma_ 0\, + \,\gamma_ 1^2\, + \,
2\,\beta_0\, \gamma_ 2 \,+\, \gamma_ 0\, \gamma_ 2)\Big) \,\eta^2
\,+\,10 \Big(12 \,\beta_0^3\, \gamma_ 1 \,+ \,\gamma_ 0^3 (3 \, \beta_ 1 \,+ \,
2 \,\gamma_ 1)\\[.1cm]
&+  \beta_0\, \gamma_ 0^2\, (13\, \beta_ 1 \,+\, 12\, \gamma_ 1) \,+ \,
\beta_0^2\, \gamma_ 0\, (13 \,\beta_ 1\, + \,22\,\gamma_ 1)\Big) \,\eta^3 
\,-\, \gamma_0\, \Big(24\, \beta_0^4 \,+\, 50\, \beta_0^3\, \gamma_ 0 \\[.1cm]
&+  35\, \beta_0^2\,\gamma_0^2 \,+\, 10\, \beta_0 \,\gamma_0^3\, + \,\gamma_0^4\Big) \,\eta^4 \Big]
\,+\, \mathcal{O}(a^6(s^*))\Bigg\}
 \end{aligned} &
\end{flalign}

where $\eta = \ln (s/s^*)$.

This is the updated series expansion of the quark mass renormalization group equation to five-loop order. Up to three-loop order Eq.(\ref{eq:A.8}) agrees exactly with \cite{kniehl1996dependence}, and up to four-loop order with \cite{chetyrkin1997estimations}.

For three active quark flavours, substituting the known values of the $\gamma_i$ and \(\beta_i\) coefficients into Eq.(\ref{eq:A.8}) results in 
\begin{flalign} 
& \begin{aligned} \label{eq:A.9}
\overline{m}_q(s) \,& = \, \overline{m}_q(s^*) \, \Bigg\{1 - a(s^*) \,\gamma_ 0 \,\eta + a^2(s^*) \Big[\frac{1}{72} \Big(-303 + 10 n_f\Big) \,\eta + \dfrac{1}{24} \Big(45 - 2 n_f\Big)\,\eta^2\Big] \\[.2cm]
&+ a^3(s^*)\Big[\Big(-\frac{1249}{64} +(\frac{277}{216} +\frac{5 \zeta_3}{6})n_f + \frac{140}{81} n_f^2\Big)\, \eta + \Big(\frac{607}{32} - \frac{233}{144} n_f + \frac{5}{216} n_f^2\Big)\, \eta^2 \\[.2cm]
& + \Big(-\frac{65}{16} + \frac{7}{18} n_f + \frac{1}{108} n_f^2 \Big)\, \eta^3\Big] 
\\[.2cm]
&+ a^4(s^*)\Big[ \Big(-98.943 + 19.108 n_f -0.276 n_f^2 -0.006 n_f^3 \Big)\, \eta
+ \Big(-146.861 -23.571 n_f 
\\[.2cm]
&-8.120 n_f ^2  + 0.432n_f^3 \Big)\, \eta^2 + \Big(-69.086 + 9.698 n_f -0.389 n_f^2 + 0.004 n_f^3 \Big)\, \eta^3 
\\[.2cm]
&+ \Big( 9.395 - 1.407 n_f + 0.070 n_f^2 -0.001 n_f ^3 \Big)\,\eta^4 \Big] 
\\[.2cm]
&+ a^5(s^*)\Big[\Big(-559.707 + 143.686 n_f -7.482 n_f^2 -0.108 n_f^3 + 0.0001 n_f^4\Big)\, \eta \\[.2cm]
&+ \Big(\dfrac{1}{2 n_f - 33} (-29836.577 + 8585.863 n_f + 22.617 n_f^2 -98.278 n_f^3 + 4.520 n_f^4 - 0.004 n_f^5)\Big)\, \eta^2 \\[.2cm]
&+ \Big(-775.076 + 164.071 n_f + 26.364 n_f^2 - 3.556 n_f^3 + 0.096 n_f^4\Big)\,\eta^3 \\[.2cm]
&+ \Big(230.956 -45.430 n_f + 3.081 n_f^2 -0.082 n_f^3 +0.0006 n_f^4\Big)\, \eta^4 \\[.2cm]
&+ \Big(-22.547 + 4.630 n_f - 0.356 n_f^2 + 0.012 n_f ^3 - 0.0002 n_f^4\Big)\, \eta ^5 \Big]
\,+\, \mathcal{O}(a^6(s^*))\Bigg\}
 \end{aligned} &
\end{flalign}

with $\zeta_n$ the Riemann zeta-function,  $n_f = 3$ in the light quark sector and $\eta = \ln (s/s^*)$.

\section{Evaluating the Accuracy of the Series Expansion of \texorpdfstring{$\overline{m}_q(s)$}{mq(s)}}

Eq.(\ref{eq:A.8}) is the perturbative series expansion of the running quark mass $\overline{m}_q(s)$ in powers of $\eta = \ln (s/s^*)$, with the initial value $\overline{m}_q(s^*)$ to five-loop order. We are now interested in the effect of the latest loop order (i.e. the $\mathcal{O}(a^5(s^*))$ term). To do this we compare Eq.(\ref{eq:A.8}) with the four-loop series determined by \cite{chetyrkin1997estimations}. 
Alternatively, we could directly numerically integrate Eqs.(\ref{eq:A.1})-(\ref{eq:A.2}) to find the running coupling $a_s(s)$ and running quark mass $\overline{m}_q(s)$ (noting that discontinuities arise at flavour thresholds). This is the method employed in RunDec, a Mathematica (and C) package used for the decoupling and running of the strong coupling constant and quark masses developed by \cite{chetyrkin2000rundec} and now in its third version.

\begin{figure}[htb]
\centering
\includegraphics[width=1\textwidth]{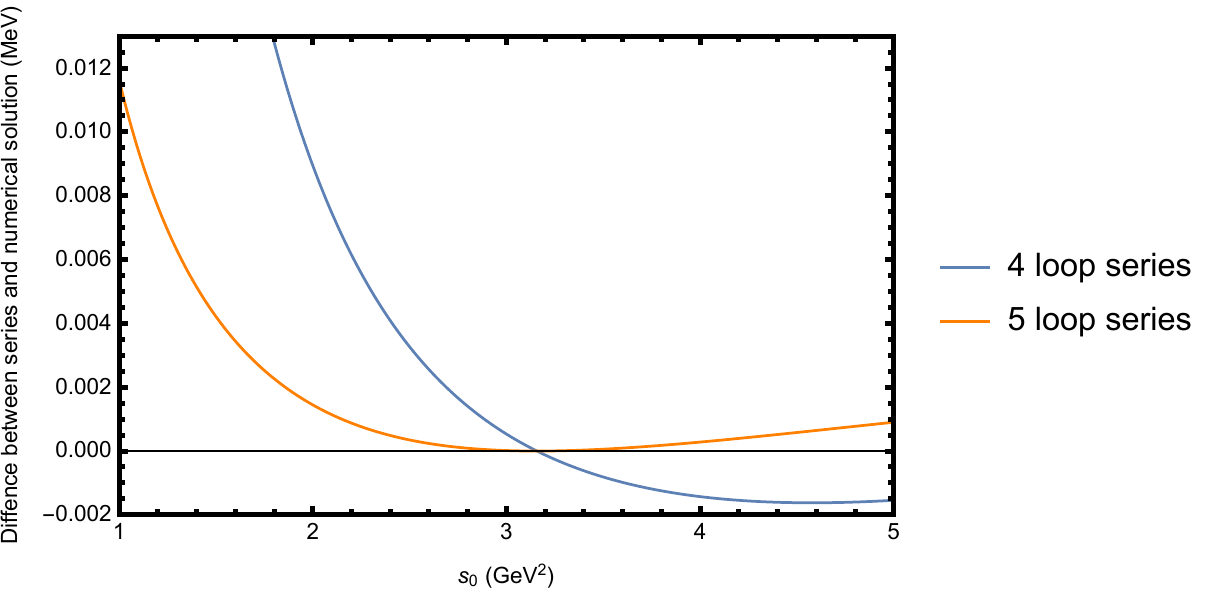}
\caption{\label{fig:difSeriesvsIntegral} The local error function, $f(s_j) \, = \, r(s_j) \, - \, k(s_j)$, where $r(s_j)$ is the reference value of $\overline{m}_{ud}(s_j)$ with a scale dependence calculated by direct numerical integration of the quark mass RG equation, and $k(s_j)$ is the value of $\overline{m}_{ud}(s_j)$ with a scale dependence as either the five-loop series expansion (orange) or the four-loop series expansion (blue).}
\end{figure}

Fig.(\ref{fig:difSeriesvsIntegral}) provides a local error analysis, by plotting the difference between the direct numerical integration of Eq.(\ref{eq:A.2}) for the running of $\overline{m}_{ud}(s)$ and i). the perturbative series solution to five-loop order (Eq.(\ref{eq:A.8})), ii). the perturbative series solution to four-loop order \citep{chetyrkin1997estimations}. Varying the energy scale between $1 \, \text{GeV}^2$ and $5 \, \text{GeV}^2$ in increments of 0.001, describes 4001 points at which to evaluate $\overline{m}_{ud}(s)$. We set the initial quark mass condition to be $\overline{m}_{ud}(s^*=(2 \, \text{GeV})^2) = (3.9 \pm 0.2 )\, \text{MeV}$ \citep{DMS2018up}, where $\overline{m}_{ud}(s)$ is defined as

\begin{equation} \label{eq:A.10}
\overline{m}_{ud} (s)\equiv \frac{\overline{m}_u(s) \, + \, \overline{m}_d(s)}{2} 
\end{equation}

We have also made use of the strong coupling constant $\alpha_s((2 \, \text{GeV})^2)= \, 0.307 \pm 0.013 $ which is found using the perturbative series expansion of the strong coupling RG equation \citep{davier2006physics} with the initial condition $\alpha_s(m_\tau^2 = 3.16 \, \text{GeV}^2) = 0.328 \pm 0.013$ \citep{pich2017precision}.

The direct numerical integration approach can be used as a reference from which we statistically compare how well it is approximated by the four-loop \citep{chetyrkin1997estimations} and by the five-loop (Eq.(\ref{eq:A.8})) series expansion. Two common statistical evaluation criteria: Root Mean Squared Error (RMSE) and Mean Absolute Error (MAE) are used to provide a global error analysis. The Root Mean Squared Error is defined as $\text{RMSE} = \sqrt{\frac{1}{n}\sum_{j=1}^{n} (r(s_j) - k(s_j))^2}$, and the Mean Absolute Error is calculated as $\text{MAE} = \frac{1}{n}\sum_{j=1}^{n} |r(s_j) - k(s_j)|$. \\[0.5mm] 
Where $r(s_j)$ is our reference i.e. $\overline{m}_{ud}(s_j)$ calculated by directly numerically integrating Eq.(\ref{eq:A.2}) at each point $j$ in the $s$ range described; and $k(s_j)$ is the quark mass $\overline{m}_{ud}(s_j)$ calculated using the perturbative series solution to either the four- or five-loop order at a particular point $j$ within the $s$ range. 
The MAE can be interpreted as the average error rate, while the RMSE is more sensitive to a large deviation between the function and the reference function at a single point. The MAE and RMSE for the four- and five-loop perturbative series solution are given in Table (\ref{tab:RGE5}). The largest absolute deviation is also given, in order to provide context for the MAE and RMSE values.

\begin{table}[htb]
\caption {Error evaluation for the forth-loop \citep{chetyrkin1997estimations} and by the fifth-loop (Eq.(\ref{eq:A.8})) series expansion, using the direct numerical integration of the quark mass RG equation as a reference} \label{tab:RGE5}
\begin{tabular}{@{}lll@{}}
\hline
\multicolumn{1}{l|}{\textbf{Statistic\,\,\,\,\,}}               & \multicolumn{1}{l|}{Five-loop series solution \,\,\,\,\,\,\,\,\,\,} & \multicolumn{1}{l}{Four-loop series solution \,\,\,\,\,\,} \\ \hline
\multicolumn{1}{l|}{\textbf{Mean Abs. Error}}         & \multicolumn{1}{l|}{0.0015}        & \multicolumn{1}{l}{0.0079}        \\ \hline
\multicolumn{1}{l|}{\textbf{Root Mean Squared Error}} & \multicolumn{1}{l|}{0.0029}        & \multicolumn{1}{l}{0.0146}        \\ \hline
\multicolumn{1}{l|}{\textbf{Largest Abs. Deviation}}  & \multicolumn{1}{l|}{0.0116}        & \multicolumn{1}{l}{0.0578}        \\ \hline
\multicolumn{1}{l|}{\textbf{Smallest Abs. Deviation}} & \multicolumn{1}{l|}{0}                 & \multicolumn{1}{l}{0}                 \\ \hline
\end{tabular}
\end{table}

From Fig.(\ref{fig:difSeriesvsIntegral}). and the low MAE and RMSE in Table (\ref{tab:RGE5}), we conclude that the five-loop perturbative series solution for the quark mass does not deviate significantly from the direct numerical integration of the mass RG equation. Hence the $\mathcal{O}(a_s^5)$ correction to the series solution of the quark mass RG equation is a valuable addition. 

\section{Validation of Results}

It has been established in Section \ref{sec:seriesexpansion} we find the peturbative series solution of the running quark mass by focusing on the separated RG equation (Eq.(\ref{eq:A.5})): integrating the given integral, using Rubi, exponentiating the result, followed by Taylor expanding around a reference point $s = s^*$. Mathematica, in comparison, can not find the peturbative solution in this way, since it fails to evaluate the integral in Eq.(\ref{eq:A.5}) into a useful form. 

This aside, we can validate the result through a different method, this time more agreeable to Mathematica. The method relies on: i). interchanging the limiting processes (performing the series expansion before integration)\footnote{To be able to replace the integrand with its Taylor expansion, and then integrate term by term; a sufficient criterion is that the series expansion converges uniformly. This is indeed the case here, with the higher order terms having a decreasing contribution.}, ii). calculating the indefinite integral, followed by using the FTC to find the definite integral, iii).  performing a second Taylor expansion after exponentiating the resultant integral.

The first stage of this process is to series expand the integrand of Eq.(\ref{eq:A.5}) which yields
\begin{flalign} 
& \begin{aligned} \label{eq:A.11}
\frac{\gamma(a)}{\beta(a)} \,& = \, \frac{\gamma _0}{a\, \beta _0} \,+\, \frac{\beta _0\, \gamma _1\,-\,\beta _1\, \gamma _0}{\beta _0^2} \, + \, a \left(\frac{\left(\beta _1^2 \,-\, \beta _0 \, \beta _2\right) \gamma _0}{\beta _0^3}-\frac{\beta _1 \, \gamma _1}{\beta _0^2}+\frac{\gamma _2}{\beta _0}\right) \\[.2cm]
&+ \, a^2 \left(\frac{\left(-\beta _1^3\,+\, 2 \,\beta _0 \,\beta _2 \,\beta _1 \,-\, \beta _0^2 \,\beta _3\right) \gamma _0}{\beta _0^4}+\frac{\left(\beta _1^2 \,-\, \beta _0 \,\beta _2\right) \gamma _1}{\beta _0^3}-\frac{\beta _1 \,\gamma _2}{\beta _0^2}+\frac{\gamma _3}{\beta _0}\right) \,+\, \mathcal{O}(a^3)
 \end{aligned} &
\end{flalign}
where the higher order terms are given in the supplementary Mathematica notebook.

Eq.(\ref{eq:A.10}) is now easily integrated with respect to $a$. After taking the definite integral and exponentiating, we then perform a second Taylor expansion in order to yield the resultant series solution. See the supplementary Mathematica notebook for further details. 

While the double series expansion may seem nonintuitive, it is able to reproduce the quark mass series expansion to five-loop order. Which, in turn, provides validation to the central method of this paper -- using Rubi.

\section{Concluding Remarks}

The case for using Rubi as a tool in this situation, and in other Science, Technology, Engineering and Mathematics (STEM) research areas, is thus: it provides a lucid and intuitive approach to solving integrals, which other CAS systems are often unable to solve directly. We have shown this through the motivating example of series solution of quark mass renormalization group equation. \\
\vspace{10mm}

\noindent{\bf Acknowledgements:}
The authors wish to thank Hubert Spiesberger for insightful discussions, and Giulia Zanderighi for a helpful comment about interchanging the process of integration and series expansion.\\

\noindent{\bf Notice:} The Mathematica$^{\circledR}$ code used is attached as a supplementary resource.\\

\clearpage
\bibliographystyle{agsm}
\bibliography{bibliography}
\end{document}